\documentclass{emulateapj}
\usepackage{times}
\usepackage{graphicx}

\slugcomment{For ApJ Letters}

\shorttitle{PSR J2215+5135 Spectroscopy}
\shortauthors{Romani et al.}

\begin{document}

\title{Keck Spectroscopy of Millisecond Pulsar J2215+5135: a Moderate-$M_{\rm NS}$, High-Inclination Binary}

\author{Roger W. Romani\altaffilmark{1}, Melissa L. Graham\altaffilmark{2}, Alexei V. Filippenko\altaffilmark{2}, and
Matthew Kerr\altaffilmark{3}} 
\altaffiltext{1}{Department of Physics, Stanford University, Stanford, CA 94305-4060,
 USA; rwr@astro.stanford.edu}
\altaffiltext{2}{Department of Astronomy, University of California, Berkeley, CA 94720-3411, USA}
\altaffiltext{3}{CSIRO Astronomy and Space Science, Australia Telescope National Facility, PO Box 76, Epping, NSW 1710, Australia}

\begin{abstract}

	We present Keck spectroscopic measurements of the millisecond pulsar 
binary J2215+5135. These data indicate a neutron-star (NS) mass $M_{\rm NS}=1.6\,{\rm M}_\odot$, much
less than previously estimated. The pulsar heats the companion face to 
$T_D\approx9000$\,K; the large heating efficiency may be mediated by the intrabinary
shock dominating the X-ray light curve. At the best-fit inclination $i=88.8^\circ$,
the pulsar should be eclipsed. We find weak evidence for such eclipses in the pulsed
gamma-rays; an improved radio ephemeris allows use of up to 5 times more {\it Fermi}-LAT
gamma-ray photons
for a definitive test of this picture. If confirmed, the gamma-ray eclipse provides
a novel probe of the dense companion wind and the pulsar magnetosphere.
\end{abstract}


\section{Introduction}

	PSR J2215+5135 is a recycled millisecond pulsar (MSP) with $P=2.6$\,ms and 
${\dot E}=7.4\,I_{45}\times10^{34}\,{\rm erg\,s^{-1}}$ (where $I_{45}$ is the 
neutron star [NS] moment of inertia in units of $10^{45}\,{\rm g\,cm^2}$), in
a $P_b=4.14$\,hr orbit with a heated low-mass stellar companion. This ``redback'' (RB)
system was discovered in 350\,MHz Green Bank Telescope (GBT) observations 
(reported by Ray et al. 2012) of an
unidentified {\it Fermi}-LAT \citep[Large Area Telescope,][]{LAT} gamma-ray source. The radio
dispersion measure ($69.2\,{\rm cm^{-3}\,pc}$) provides a distance estimate $d\approx3$\,kpc, and
this relatively close and bright system can be subject to detailed study.

	Schroeder \& Halpern (2014, hereafter SH14) obtained high-quality $BVR$ light curves
of the companion over many orbits. It varies from $V\approx18.7$ to 20.2\,mag and
thus is well measured with modest-aperture telescopes. Fitting these light curves
with the ELC code \citep{oh00} and a photometry table generated from the PHOENIX model atmospheres
\citep{huet13}, SH14 inferred a low system inclination and large pulsar mass, $M_1=2.45^{+0.22}_{-0.11}\, {\rm M}_\odot$.
This is potentially very important, since it would be one of the highest-known NS masses.
Such objects can provide strong constraints on the equation of state at supernuclear 
densities \citep{lp07,slb13}.

	However, their fit showed several peculiarities. The unheated ``night'' face of the
companion was found to have $T_c=3790^{+35}_{-25}$\,K and the heated ``day'' side $T_D=4899^{+34}_{-23}\,$K in the
ELC fits, in poor agreement with the observed bluer colors. The light curve seemed shifted 
by $\Delta\phi\approx-0.01$ with respect to the radio-pulse ephemeris. SH14 suggest that the blue colors
could be a hot quiescent disk, while the phase shift may be an effect of an intrabinary
shock. Also, \citet{genet14} observed the system in the X-rays with CXO, finding an
X-ray minimum near orbital phase $\phi\approx0.25$ (pulsar superior conjunction, radio eclipse), 
which they interpret as due to variable obscuration of emission from an intrabinary shock 
around the companion.

	To probe these peculiarities, we have obtained Keck spectra across the binary orbit. 
Our data show no evidence for disk or shock emission lines, but imply much higher system 
inclination, with higher companion temperatures and
stronger pulsar heating. This lowers the inferred pulsar mass, but raises the possibility that
the pulsed magnetospheric emission is eclipsed by the companion star. 
Gamma-ray data currently show limited evidence for such eclipse, but additional observations can 
test this hypothesis. 

\section{Keck Spectroscopy}

	The system was observed on three occasions. In each case we aligned the
long slit at position angle $76^\circ$ (N through E) to include two bracketing stars 
for monitoring purposes (see the finder chart in Breton et al. 2013). The $R=17.6$\,mag star
at USNO-B1 J2000 position $\alpha=22^{\rm hr}15^{\rm m}32.031^{\rm s}$, $\delta=+51^\circ35'35.46''$ (C1) 
appears to show radial-velocity (RV) variations between epochs, 
so we use the $R=18.1$\,mag star $8^{\prime\prime}$ away at J2000 position 
$\alpha = 22^{\rm hr}15^{\rm m}33.498^{\rm s}$, $\delta=+51^\circ35'38.88''$ (C2)
as our reference object.

First, using the DEep Imaging Multi-Object Spectrograph (DEIMOS; Faber et al. 2003) on Keck-II, 
we obtained $7\times600$\,s exposures covering 4450--9060\,\AA\ on 2014 October 2 (UT dates
are used throughout this paper; MJD 56932) with
$\sim4.7\,$\AA\ resolution through a $1^{\prime\prime}$-wide slit. 
These data covered $\phi_B= 0.685$--1.025, where $\phi_B=0.75$ is optical maximum (``Noon'')
and the pulsar ascending node is at $\phi_B=0$.  The radio ephemeris defines
$T_{\rm ASC}=55186.164485831$\,MJD and $P_B=0.172502104907$\,d. 
For improved blue coverage, we observed again on 2014 October 24 
(MJD 56954) by taking $6\times 600$\,s exposures with the Low Resolution Imaging Spectrometer 
(LRIS; Oke et al. 1995) on Keck-I using the $1^{\prime\prime}$ slit and
the 5600\,\AA\ dichroic splitter. The blue-camera 600/4000 grism
provided $\sim4$\,\AA\ resolution, while the red-camera 
400/8500 grating delivered $\sim7$\,\AA\ resolution. These data covered 
wavelengths $\sim 3200$--10,260\,\AA\ and orbital phases $0.501<\phi_B<0.723$.
Finally, we observed with the same setup on 2014 November 20 (MJD 56981), covering
orbital phase $0.202<\phi_B<0.381$ by taking $5\times600$\,s exposures. Calibration spectra
of blue (BD$+28^\circ4211$) and red (BD$+17^\circ4708$) spectrophotometric standards were 
obtained during each run. The spectra
were subject to standard IRAF reductions, including optimal extraction.

	Figure 1 shows the reduced, calibrated LRIS spectra at three orbital phases. 
The companion is significantly hotter than the SH14 ELC solution, with an effective spectral class
A2\,V near optical maximum brightness ($T_{\rm eff}\approx8970$\,K) and G0\,V near minimum ($T_{\rm eff}\approx6030$\,K),
as determined by line-ratio comparison with spectral standards.
No strong emission lines or nonthermal component are seen, although a weak variable 
blue continuum peaking at $\phi_B\approx0.75$ is not at present excluded.

\begin{figure}[t!!]
\vskip 8.3truecm
\includegraphics{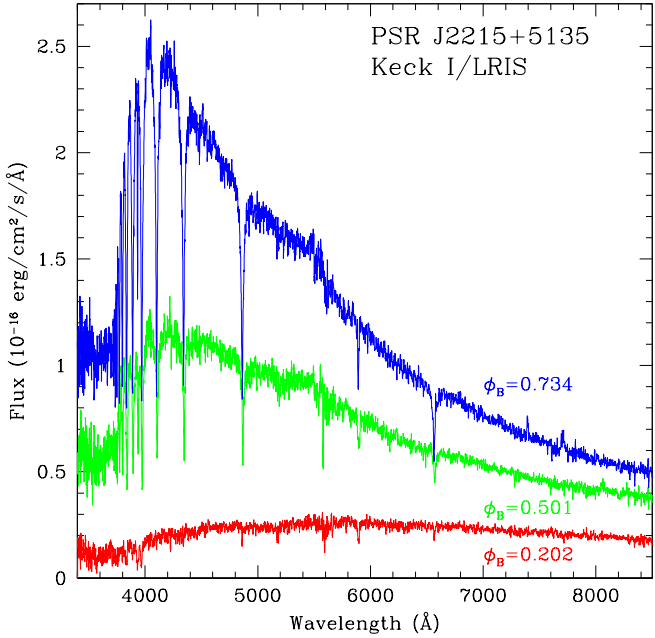}
\begin{center}
\caption{\label{spectra} 
Keck-I/LRIS spectra of J2215+5135, showing maximum brightness (``Day''; $\phi_B=0.73$),
``Dawn'' ($\phi_B=0.50$), and ``Night'' ($\phi_B=0.20$). Each spectrum is a single 600\,s exposure.
Broad features near 5500\,\AA\ are artifacts from incomplete correction for the dichroic response.
}
\end{center}
\vskip -0.5truecm
\end{figure}

\begin{figure}[t!!]
\vskip 8.3truecm
\includegraphics{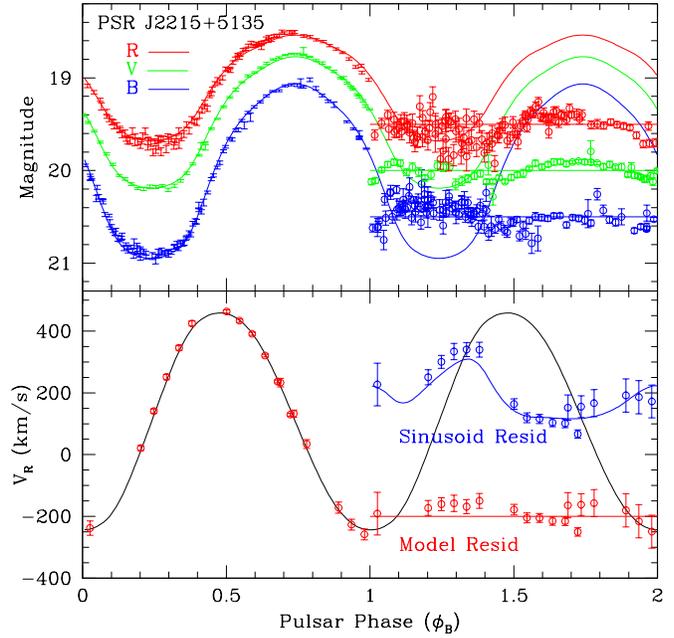}
\begin{center}
\caption{\label{LC/RV} 
Top: $BVR$ MDM photometry \citep{sh14}.
Bottom: Keck DEIMOS/LRIS radial velocities. Two periods are shown, with data and the large-$i$ 
(``HiT'')
ELC model in the first cycle.  The second period shows the model residuals, multiplied
by a factor of 3. In the velocity panel, we also show the residuals after a best-fit sinusoid is removed from both 
the model and the data.
}
\end{center}
\vskip -0.5truecm
\end{figure}

	The spectra show a wealth of stellar absorption features, and so we 
measure the RV variation using the IRAF XCASO script \citep{km98}.
The templates were drawn from the Indo-US library of Coud\'{e} Feed spectra \citep{vet04},
selecting dwarf stars with known RV from a range of well-determined
spectral classes. The cross-correlation statistic was $R\approx30$--60 
for the day-phase October 24 LRIS data and $R\approx20$--25 for the night-phase 
November 20 spectra. With the more-limited blue coverage on October 2, the
cross correlations were weaker, giving $R\approx10$--20 and larger RV uncertainties.
Since C1 shows $\sim 50\,{\rm km\,s^{-1}}$ RV variation between
epochs, we relied on C2 for our RV tie. This star matched best to a K0\,V spectrum.
Cross correlation delivered velocity uncertainties of $\sim 2.5\, {\rm km\,s^{-1}}$
(LRIS) and $\sim 3.5$--4.0\, ${\rm km\,s^{-1}}$ (DEIMOS). To remove residual drift in the 
wavelength solution, we corrected the pulsar companion velocities to bring the
measured C2 velocities into agreement. The C2 measurement errors, plus systematic 
$3\, {\rm km\,s^{-1}}$ errors estimated from the variance between cross correlations with
different templates, were added in quadrature to the individual RV uncertainties 
to produce the final RV uncertainty.
These RVs and errors are plotted in Figure \ref{LC/RV} as a function of orbital phase computed from 
the barycentered times of the exposure midpoints.
Lacking velocity-standard observations spanning the companion spectral class during each run,
we may have additional systematic uncertainty in the absolute RV.

\section{Light-Curve Comparison and Modeling Constraints}

	To compare with the light-curve data, we plot the phased $BVR$ magnitudes and 
uncertainties published in Figure 2 of SH14; we ignored upper-limit points (generally
from epochs of poor photometry), extracting 103 $B$, 55 $V$, and 113 $R$ magnitudes.
These data and our new RV measurements were analyzed with two popular
light-curve modeling programs: the ELC code and the ICARUS code (Breton et al. 2011).
For both we can run in ``MSP'' mode, with the projected pulsar orbit 
$x_1=a_1\, {\rm sin}\,i = 0.468141433$\,lt-s from the radio ephemeris.
Both codes model pulsar heating of the companion as illumination by a point (X-ray) source,
generating filter-specific light-curve models and estimating binary-system parameters
via model fits. The temperature model is determined by the underlying
temperature of the star (actually $T_c$ of the unheated ``night''
face), and a heating flux denoted $L_H$.  The observed light curves are then
sensitive to the orbital inclination $i$ and the mass ratio $q=M_{\rm NS}/M_c$.
The heating power can be related to the effective temperature of the heated face
$T_D$ by
$$
L_H = (T_D^4-T_N^4) 4\pi [x_1(1+q)]^2 \sigma/{\rm sin^2}i
\eqno(1)
$$
(effective albedo = 0).
The Roche lobe filling factor $f_c$ is also relevant; as concluded by SH14, ELC fits
require $f_c\approx1$. It is also large ($f_c>0.86$) in 
the ICARUS fits. SH14 find that the heating center is phase shifted.
We concur, but find $\Delta \phi =-0.0089$, smaller than the 
$\Delta \phi =-0.0140\pm 0.0005$ of their Table 2.

	The two codes have some model differences. For the ELC code,
one can simultaneously fit the observed RV points, giving 
additional sensitivity to $q$ and direct estimates of the component masses. 
However, the code fits normalized light curves for each band separately, ignoring 
the instantaneous colors and observed magnitude. When searching high-quality data
for the minimum near a correct physical model, this normalization should not affect
the fit values and insulates the results from systematic interband photometry errors.
The ICARUS code, in contrast, fits the observed magnitudes directly, using
the instantaneous colors and providing estimates of the distance modulus
and the extinction.

\begin{figure}[b!!]
\vskip 8.8truecm
\includegraphics{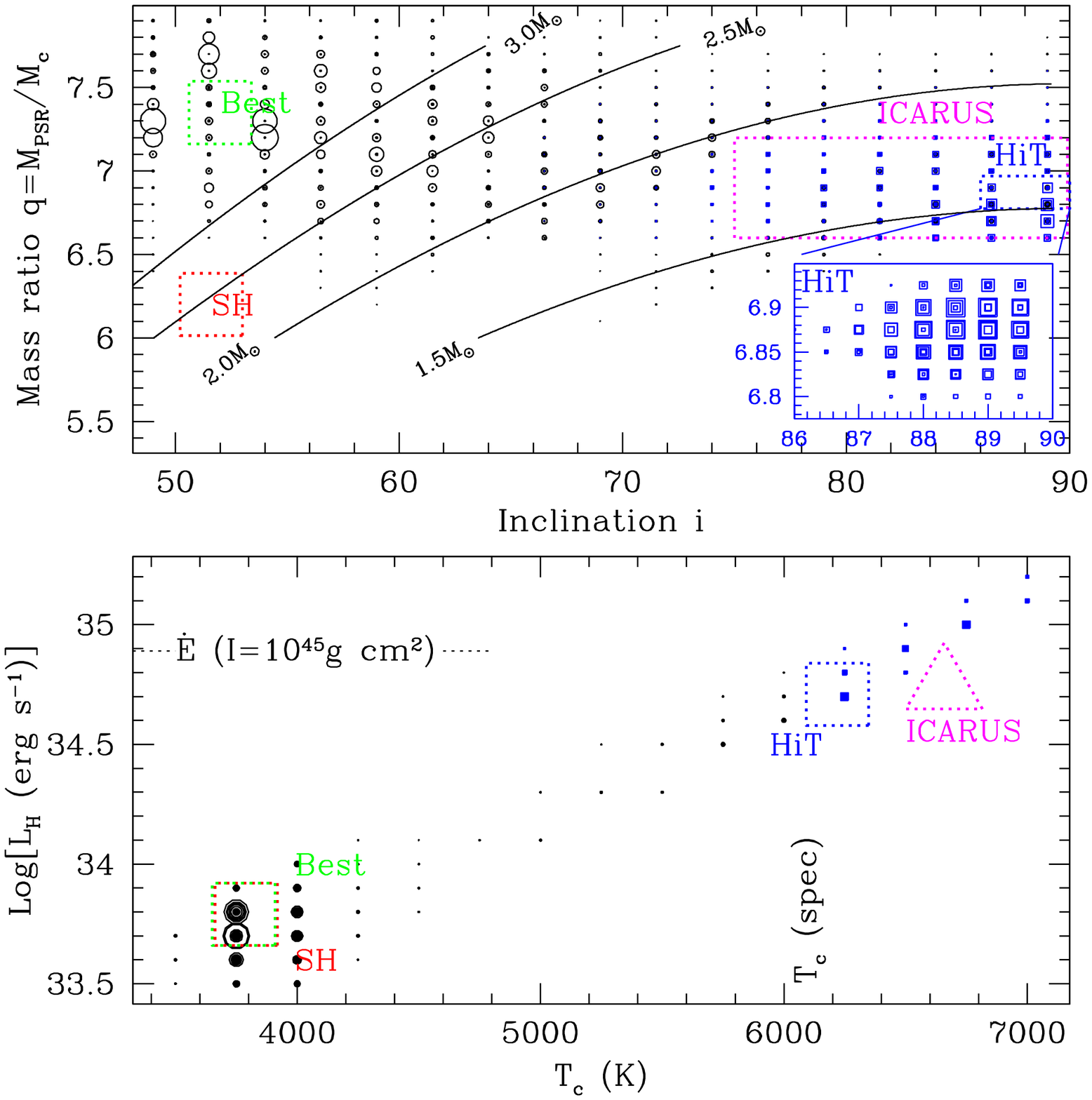}
\begin{center}
\caption{\label{Fitplot} 
Four-parameter ELC photometry/RV fit $\chi^2$ projected onto two planes; large dots indicate low $\chi^2$.
Square dots are for $T_c>6000$\,K, consistent with the spectroscopic classification. Dotted
boxes indicate locations of the various fits in Table \ref{Fits} (measured from finer grids). 
The ``HiT'' ELC-fit minimum consistent with the spectroscopy is shown
in the inset box of the upper panel; here the point size scales inversely with $\chi^2$, decreasing
to 0 at $\Delta \chi^2/\nu=3.0$.  Upper panel lines show the inferred pulsar mass.
}
\end{center}
\vskip -0.8truecm
\end{figure}

	For the ELC modeling, we wish to compare with the SH14 results, and so we used
the same color table (kindly shared by J. Schroeder) generated from the PHOENIX 
atmosphere models \citep{huet13} with an extension to $T_{\rm eff}>$ 10,000\,K from the ATLAS9 
atmospheres \citep{ck04}. For the ICARUS code we collected Harris $BVR$ color tables from
the PHOENIX models tabulated at the Spanish Virtual Observatory (svo2.cab.inta-csis.es).
Our results are summarized in Table \ref{Fits} and Figure \ref{Fitplot}. 

	Our ELC fit to the $BVR$ magnitudes finds parameters similar to those of SH14,
except that their solution lies to the side of a broad minimum; the global
minimum lies at considerably higher $q\approx7.5$, with $\chi^2/\nu=3.25$ (the minimum
in their estimated $1\sigma$ contours is $\chi^2/\nu =3.27$). These $\chi^2/\nu$ values are about twice
those reported by SH14, but the model light curves and residual plots look quite similar to 
their results across this range of $q$.  These ELC photometry fits give much lower temperatures 
and heating powers than allowed by our Keck spectra.  The ICARUS photometry fit gives a 
dramatically different picture, finding $T_c\approx6600$\,K, much more consistent with the 
spectral estimate. $L_H$ is also much larger. The fit does not strongly constrain $q$.
The best fits prefer large extinctions; we fix $A_V=0.62$\,mag, the maximum in 
this direction from the \citet{sf11} models. The distance modulus is 12.98\,mag or $d\approx3.9$\,kpc,
not inconsistent with the dispersion-measure estimate.

We next explored ELC fits of the combined photometry and spectroscopy RV datasets. Since 
there are only 18 velocity points,
these could be overwhelmed by the photometric data. Accordingly, we fit using the RV points
5 times each, effectively increasing the weights to compete with the photometry sets.
Testing with only 1 and as many as 10 RV sets revealed that the fit results are 
insensitive to this choice, but the fitting converged more rapidly to the minima with the increased RV weights.

\begin{deluxetable}{llrrrrr}[t!!]
\tablecaption{\label{Fits} Binary Model Fits}
\tablehead{
\colhead{Param.} & \colhead{SH14}& \colhead{P\tablenotemark{a}} & \colhead{P/S\tablenotemark{b}} 
& \colhead{HiT\tablenotemark{c}} & \colhead{ICARUS} 
}
\startdata
$i$         &$51.6^{+2.7}_{-2.1}$  & 51.8 & 52.0 & 88.8 &75--90\cr
$q$         &$6.20\pm0.25$          & 7.50 & 7.35& 6.89 &{\it (0.87--0.99)}\tablenotemark{d}\cr
$T_c$ (K)   &$3790^{+35}_{-25}$    & 3780 & 3780& 6220 &$6637\pm12$ \cr
Log($L_H$)   &33.79                   & 33.79&  33.77& 34.71& 34.76\cr
$\chi^2/\nu$&[3.29]\tablenotemark{e}&3.25 &  2.98& 4.36 & 5.05\cr
$M_{\rm NS}$&$2.45^{+0.22}_{-0.11}$& 4.14&  3.88& 1.59 & --  \cr
$M_c$ &$0.396\pm 0.045$      & 0.55&  0.53& 0.23 & -- \cr
\enddata
\tablenotetext{a}{P: ELC photometric fit}
\tablenotetext{b}{ P/S:ELC photometric/spectral fit (``Best'' in Figure 3.)}
\tablenotetext{c}{HiT ELC photometric/spectral fit w/ $T_c>6000$\,K constraint.}
\tablenotetext{d}{Roche fill factor. $f_c$ increases with $i$; all $f_c>0.9$ imply $i\approx 90^\circ$.}
\tablenotetext{e}{Our fit value; SH14 report $\chi^2/\nu\approx 1.5$.}
\end{deluxetable}

	Figure 3 illustrates the results from our combined photometry/RV fits, with two
panels showing the projected $\chi^2$ from a grid of models in four fit parameters. The points
are scaled so that large sizes correspond to lower $\chi^2$. With a minimum $\chi^2 \gg \nu$,
we must conclude that the models are inadequate (or the error bars are underestimated).
Thus, no particular fit value or formal uncertainty should be believed. Nevertheless, the global picture
of a pulsar-heated companion is surely correct, and the distribution of best (albeit inadequate)
fits in parameter space gives a guide to future precise solutions.

	First, we note the strong correlation of the ``good'' solutions in the $T_c$--$L_H$ 
plane. Recalling that ELC fits the normalized light curves, we can understand that this
simply preserves the height of the light-curve peak; with higher $T_c$, larger $L_H$
is needed for the same maximum. The best light-curve fits
are indeed for low $T_c$ solutions, as found by SH14. However there is a shallower minimum
at $6000\,{\rm K}< T_c <7000\,{\rm K}$, consistent with the colors and spectroscopic temperatures. This
local minimum is not evident when fitting only photometry. The RV data require $q$ substantially
larger than in the pure photometry fits, but $i$ remains poorly constrained,
with a wide swath of moderate $\chi^2$ displaying many local minima. The small SH14 $q=6.2$ value
is excluded, but the global minimum $i\approx52^\circ$ is quite similar to the best-fit photometry-only 
solution.  This solution has an unphysically large pulsar mass $M_{\rm NS} = 3.88\,{\rm M}_\odot$.  

	However, if we restrict attention 
to solutions with $T_c>6000$\,K, consistent with the spectroscopic classification (square dots), 
these form a minimum at large inclination $i$, and $q\approx6.9$. This shallower but
well-defined minimum, shown in detail in the upper-panel inset of Figure 3, has a much more modest 
$M_{\rm NS} = 1.59\,{\rm M}_\odot$. In Figure 2, we show the light-curve and RV models
for this solution, and the fit residuals. Note the systematic residuals over optical maximum,
especially in $V$ and $R$. This indicates an incorrect heating model. The correct heating pattern
(likely mediated by an intrabinary shock; see below) should have a surface temperature distribution
allowing a better match to the light-curve peak. The RV residuals show similar systematic departures. 
We also plot the RV points and model after the same best-fit sinusoid has been
removed. This shows that the basic nonsinusoidal terms in the center-of-light velocity are
well detected in our spectra.

This solution has very large $i$, raising the interesting possibility of a pulsar eclipse.

\begin{figure}[t!!]
\vskip 5.2truecm
\includegraphics{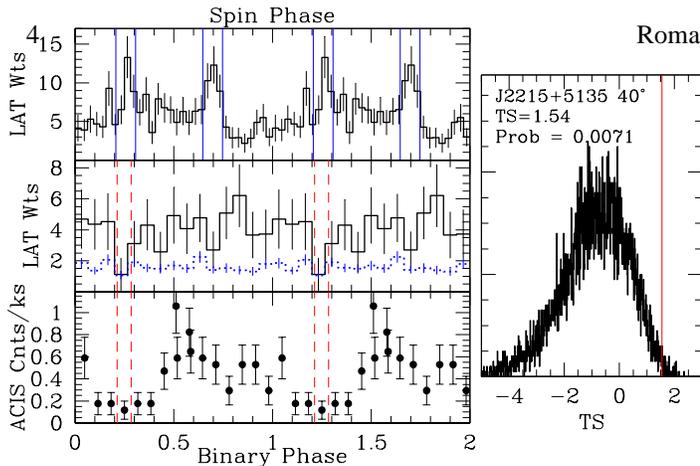}
\begin{center}
\caption{\label{LCEcl} 
Left: Top to bottom: gamma-ray pulse profile (with peak phase bins marked), orbital light curve
for the peak phase bins (solid line) and full phase (dashed line, summed weights/8), and orbital 0.3--10\,keV
light curve, showing the broad minimum. The direct secondary eclipse phase is marked by the dashed
lines, and error bars on the LAT-weighted count curves are computed as $\sigma=(1 +\sum_i w_i^2)^{1/2}$
(see 2PC).
Right: Likelihood ratio test of the weights in an orbital eclipse. The red line marks the measured TS,
with $10^4$ random realizations for $\theta_e=40^\circ$ (black histogram). Moderate significance is seen
for eclipse widths 25$^\circ$--45$^\circ$.
}
\end{center}
\vskip -0.5truecm
\end{figure}

\section{The High-Energy Orbital Light Curve}

	At our large $i$ (``HiT'') fit minimum, we expect the companion to eclipse the pulsar for 
$\Delta \phi_E = 25.4^\circ$.  The radio-pulse emission is undetected for over half 
the orbit.  However, the ionized companion winds of RB and black-widow pulsars generally 
prevent radio-pulse detection for sight lines
passing well outside the companion Roche lobe, so this alone does not indicate true eclipse.

	 We have re-examined the 17.0\,ks {\it CXO} ACIS exposure of PSR J2215+5135 (ObsID = 12466),
starting on MJD 55697.1821 and covering 1.14 orbits. As noted  by \citet{genet14}, there is a clear
shallow modulation with a minimum at pulsar superior conjunction $\phi_B\approx 0.25$ (Figure 4).
This shallow modulation might be best interpreted as variable viewing of an X-ray emitting 
intrabinary shock.

	A powerful intrabinary shock can also produce GeV LAT photons \citep{xw15}. However,
since an appreciable fraction of the object's gamma-ray flux is pulsed, we can select the magnetospheric
component and search for a true pulsar eclipse. Unfortunately, as for many RB MSPs, the wind obscuration
and $P_B$ fluctuations make radio timing very difficult. At present, the best ephemeris
available is that of \citet[2PC]{2pc}, valid MJD 55346--55911, about a fifth of the LAT mission to date. 
We selected 10,951 $E=0.1$--30\,GeV Pass 7 reprocessed `Source class' LAT photons 
within this date window and within $2^\circ$ of the pulsar. We compute weights $w_i$, the probability of being 
pulsar photons, with the LAT tool {\it gtsrcprob} using the \cite{3FGL} source spectrum parameters 
and the local background model; the mean source probability was $\langle w_i \rangle =0.037$. 
The spin-phase $\phi_S$ weighted light curve shows two peaks consistent with 
2PC results ($\phi_S=0.257$ and $\phi_S=0.697$ relative to the radio-pulse peak). If we extrapolate
this ephemeris to the full LAT dataset, the peaks are lost. We select $\Delta \phi_S=0.1$ windows
centered on the 2PC pulse peaks, to obtain events most dominated by
magnetospheric emission. Other phases represent pulsed bridge flux, unpulsed flux from the 
intrabinary shock (if any), and unmodeled background emission. The middle panel shows the 
weighted binary light curve from the peak phase window; intriguingly, the minimum bin is at 
$\phi_B=0.25$, pulsar superior conjunction. The full spin-phase light curve shows no
strong eclipse. This suggests an additional uneclipsed gamma-ray component.

	Binning clearly affects the minimum's appearance, so we desire an unbinned test
for the eclipse significance. For an eclipse width $\theta_e$, we form the test statistic
$${\rm TS} = 
\Sigma_1 {\rm log}\left [ 1+w_i\theta_e/(1-\theta_e) \right ] 
		  + \Sigma_2 {\rm log}(1-w_i),
\eqno(2)
$$
where the first sum is over the uneclipsed window and the second is over the eclipse window.
Exposure variations are already very small ($\sim 2$\% in the binned light curve), but the 
likelihood ratio test used here is insensitive to even these, since the variations are common 
to both the null and eclipse scenarios.
To estimate the probability distribution of this statistic, we scrambled weights among
the pulse peak photons and recomputed TS 10,000 times. At $\theta_e=25.4^\circ$,
we find a chance probability of 0.080 ($\sim 1.4\sigma$). However, the companion wind and photon
opacity may widen the gamma-ray eclipse. Indeed, somewhat wider eclipses, to 
$\theta_e\approx45^\circ$, show larger significance.  The best detection is at 
$\theta_e\approx40^\circ$, illustrated in the histogram at right, with a 0.7\% ($\sim 2.45\sigma$, single trial) 
probability of a false-positive detection. This is intriguing, but hardly definitive. However,
an improved radio ephemeris should provide a factor of $>4$
more photons and better pulse phase isolation, allowing a
sensitive test for a magnetospheric (pulse phased) eclipse.

\section{Conclusions}

	Our spectroscopic study of the RB MSP J2215+5135 does not support the inference, based 
on companion light-curve-shape fits, that it is a particularly massive pulsar.
Instead, the results suggest a modest $\sim 1.6\,{\rm M}_\odot$ NS mass. However, the present
fit is based on a clearly inadequate heating model assuming direct (X-ray) pulsar illumination.
Of course, for all viable ELC model fits the
heating flux is much larger than the $L_X \approx 10^{32}\,{\rm erg\, s^{-1}}$ of the {\it CXO} observation.
Indeed, as found for PSR J1311--3430 (Romani, Filippenko, \& Cenko 2015), we find that the heating 
power into the companion solid angle
is a large fraction of the spin-down flux (Figure \ref{Fitplot}, bottom); 
we argue that this indicates indirect heating by emission reprocessed in an intrabinary shock. 

With the
present direct heating model, the fits are inadequate and the critical binary inclination $i$ is 
not robustly determined. Accordingly, the fit masses are indicative only, and the statistical errors
are not useful.  There is a clear lesson here: with an inappropriate physical model, simple ELC fits
to only the light-curve shape can deliver fit minima far from the true solution. 
Using the observed colors, as in ICARUS, certainly
helps, but spectroscopic constraints are also essential. Of course, amendment of the physical model
should greatly improve the fit quality and may allow photometry-only solutions with high-quality data.

	Our temperature-constrained fits do suggest that the system is
viewed nearly edge-on. If so, this presents the interesting possibility that the pulsar magnetosphere
will be eclipsed by the companion.  Limited by the radio ephemeris
available, present evidence for such eclipses is suggestive, but not definitive. However, additional 
radio data and further analysis of the {\it Fermi} LAT photons can produce a strong test of the eclipse
hypothesis. If well measured, the width and spectrum of the pulsed flux eclipse can provide an important
new tomographic probe of the MSP magnetosphere and of the evaporative companion wind.
\bigskip

We thank J. Schroeder for comments about ELC fitting and for sharing the
PHOENIX table, WeiKang Zheng for assistance with the Keck observations, and Paul Ray for a careful
review of the manuscript.
This work was supported in part by NASA grant NNX11AO44G.  A.V.F. and M.L.G. were
supported by Gary and Cynthia Bengier, the Christopher
R. Redlich Fund, the TABASGO Foundation, and NSF grant AST-1211916.
Some of the data presented herein were obtained at the
W. M. Keck Observatory, which is operated as a scientific partnership
among the California Institute of Technology, the University of
California, and NASA; the Observatory was made possible by the
generous financial support of the W. M. Keck Foundation.

The \textit{Fermi}-LAT Collaboration acknowledges support for LAT development, operation, and data analysis from NASA and DOE (United States), CEA/Irfu and IN2P3/CNRS (France), ASI and INFN (Italy), MEXT, KEK, and JAXA (Japan), and the K.A.~Wallenberg Foundation, the Swedish Research Council, and the National Space Board (Sweden). Science analysis support in the operations phase from INAF (Italy) and CNES (France) is also gratefully acknowledged.

\end{document}